\documentclass[aps,prd,email,preprint,showpacs,showkeys,preprintnumbers,amsmath,amssymb,nofootinbib]{revtex4-1}

\usepackage{amsmath}
\usepackage{latexsym}
\def\[{\left\lbrack}
\def\]{\right\rbrack}

\def\({\left(}
\def\){\right)}

\newcommand{\be}{\begin{equation}}
\newcommand{\ee}{\end{equation}}
\newcommand{\ea}{\end{eqnarray}}
\newcommand{\ba}{\begin{eqnarray}}

\begin{document}

%\title{Fractional Dirac Bracket on Riemman Liouville Context}
%\title{\Large{Equation of state and energy conditions \\ for noncommutative wormholes}}
\title{\Large{Noncommutative wormholes and the energy conditions}}

\author{Everton M. C. Abreu$^{a,b,c}$}
\email{evertonabreu@ufrrj.br}
\author{N\'elio Sasaki$^b$}
\email{sasaki@fisica.ufjf.br}

\affiliation{${}^{a}$Grupo de F\' isica Te\'orica e Matem\'atica F\' isica, Departamento de F\'{\i}sica,
Universidade Federal Rural do Rio de Janeiro\\
BR 465-07, 23890-971, Serop\'edica, Rio de Janeiro, Brazil\\
${}^{b}$Departamento de F\'{\i}sica, ICE, Universidade Federal de Juiz de Fora,\\
36036-330, Juiz de Fora, MG, Brazil\\
${}^{c}$LAFEX, Centro Brasileiro de Pesquisas F\' isicas (CBPF), Rua Xavier Sigaud 150,\\
Urca, 22290-180, RJ, Brazil\\
\bigskip
\today}


\begin{abstract}
\noindent It is a very well known fact that the energy conditions concerning Lorentzian wormhole (WH) solutions of Einstein equations are violated.  Consequently, attempts to avoid the violation of the energy conditions constitutes one of the main areas of research in WH physics.  On the other hand, the current literature show us that noncommutativity is one of the candidates to understand the physics of the early Universe.  In this letter we show that the null and weak energy conditions violations do not happen when the WH background geometry is a noncommutative one.  We also construct the shape function and an equation of state for this noncommutative WH.
\end{abstract}

\maketitle

\pagestyle{myheadings}
\markright{\it Noncommutative wormholes and the energy conditions}

\newpage

%\setlength{\baselineskip} {20 pt}

%\section{Introduction} 

Dwelling in the imaginary of science fiction authors, a wormhole (WH) play a fundamental role when they talk about time travel.  Its popularity came from theoretical solutions of Einstein equations which show that WHs can act as tunnels linking two distant places \cite{basics} (for a recent review see \cite{basics2}).   Hypothetically, traversable WHs would permit effective superluminal travels.  In this case however, the speed of light is not locally surpassed \cite{1}, of course.

%A WH can be classified in two different categories, traversable and non-traversable.  
A traversable WH  permits the passage of a traveler since the throat is not closed.  For this property to be possible it is considered that the WHs would be formed by a kind of {\it exotic matter}, which violates the weak energy condition (WEC), and would violate the gravitational attraction (\cite{basics2} and references therein).  Therefore WHs have been viewed as quite different from black holes (BH).  Topologically speaking we can say that traversable WHs are short ``handles" in spacetime topology that links, as we said above, widely separated regions of the same Universe.  Or  ``bridges," which links  two different spacetimes (different Universes).  In both cases we have non-trivial topologies ($R_1 \times R_2 \times S^2$) of multiply connected Universes.  Matter with negative energy density is well known in effects like the Casimir one.

To avoid singularities and horizons, one must construct the throat with nonzero energy-momentum tensor.  In \cite{sh} the authors showed that under perturbation, there is a possibility that the WH collapse to a BH and the throat closes.  It is important to say that, the process of WH creation, together with extremely large spacetime curvature can be both ruled by quantum gravity conditions \cite{basics2}.  Considering the process of WH formation, we assume that some field fluctuation collapsed and consequently it originates a rotating scalar field framework \cite{mn}.  This last one has one interior region, where the rotation is not zero, and two other exterior regions, one on each side of the throat, where the rotation is zero.  The internal rotation is supposed to keep the throat stable.  And the interior field is the source of the WH.  We will see that this rotation is present in the WH under the NC regime and it is sufficient to maintain the throat open.

There are various approaches dealing with WHs, both static \cite{2} and evolving relativistic forms \cite{3}.  Both consider static WH spacetimes sustained by a single fluid object that requires the violation of the null (NEC) and WEC.  Violation of NEC on or near of a traversable WH's throat is a general WH's property. The NEC is the weakest one.  Namely, if NEC is violated, all pointwise energy conditions will be violated too.

Concerning the violation of NEC and WEC together with the property that keeps itself (the throat) open in order to permit traveling, we can say that the WHs must be formed by a (exotic)  matter, which is characterized by an energy-momentum tensor which violates the WEC.  Hence, the energy density must be negative in the framework of reference of at least a few observers.  It is important to say that although classical kinds of matter obey the WEC, quantum fields can generate negative energy densities (locally).  These last may be quite large at a given point.  
%We will talk more about this energy issue in a moment.

Hence, to analyze the WH's energy condition we have to verify the value of its energy-momentum tensor components.  After that, it was found that these traversable WHs have an energy-momentum tensor that violates the NEC.  As a matter of fact, it was found that they violate all the known pointwise energy conditions and averaged energy conditions.  And they are very important to the singularity theorems and in classical BH thermodynamics.  
%As the NEC is the weakest condition, its violation indicates that the other conditions were also broken.

Recent cosmological observations has been shown that the cosmological fluid violates the strong energy condition (SEC).  And imply, perhaps, that the NEC can be possibly violated in a classical scenario.  Under these results one can conclude that the WEC, NEC and the other energy conditions, cannot be considered underlying laws with consequences in the WHs construction.  However, one can impose frameworks that obey the energy conditions.  For instance,  in \cite{3} it was shown that if we introduce a temporal function inside the WH metric, there will be constraints on this function that provide the WH to obey the energy conditions.  This situation is critical since it has been shown that classical systems violate all the energy conditions \cite{4}.   Besides, recent cosmological observations show us that possibly the cosmological fluid violates SEC and that NEC might possibly be violated in a classical behavior \cite{7}.  Because of these reasons, one of the main areas in WH investigation, as we said before, is to study how to avoid strongly the violation of NEC \cite{basics2}.  However, NEC and ANEC are always violated for WHs spacetimes.
Therefore, we consider that an alternative structure through which WHs  naturally do not violate the energy conditions must be cherished.  

It can be shown that noncommutativity \cite{6} can eliminate the divergences that dwell in general relativity \cite{6.1}.  
%In few words, to construct a noncommutative WH (NCWH) means to change the WH point-like structure in favor of a smeared geometry in flat spacetime.  
Following this NC path, we will see that the NC features will bring new properties to WH physics.  There are other versions of noncommutative WHs (NCWHs) in the literature \cite{6.2} but they use Nicolini {\it et al} approach \cite{6.3} which is based on coherent states, differently from us.  We will introduce noncommutativity via Bopp shift used by Chaichian {\it et al} in \cite{chaichian}.

In this letter we will show that these energy violations do not happen when the WH background geometry is NC.  Namely, the NCWH energy and flaring-out conditions are satisfied.  Besides, we will construct a NC shape function and an equation of state for a NCWH.

%\section{}

We begin by writing the WH metric in the static, spherical and symmetric form with the coordinates $(t,r,\theta,\phi)$ as
\be
\label{1}
ds^2\,=\,-\,e^{2\Phi(r)}\,dt^2\,+\,\frac{1}{1-\frac{b(r)}{r}}dr^2\,+\,r^2\,(d\theta^2\,+\,sen^2\,\theta\,d\phi^2)
\ee
where $r\in (-\infty,+\infty)$ and we will assume that the redshift function $\Phi(r)=\;$constant to simplify our problem.  Eq. (\ref{1}) describes a WH geometry formed by two identical regions joined together at the throat.

The Einstein field equations are given by $G_{\mu\nu}=8\pi T_{\mu\nu}$ and we will use that $c=G=1$ to obtain the following quantities
\be
\label{2} 
\rho(r)\,=\,\frac{b'}{8\pi r^2}
\ee
\be
\label{3}
p(r)\,=\,-\,\frac{1}{8\pi}\,\frac{b}{r^3} 
\ee
\be
\label{4}
p_{tr}\,=\,\frac{1}{8\pi}\(1-\frac{b}{r}\) \[ \frac{-b' r\,+\,b}{2r^2(r-b)}\] 
\ee
which are the energy density, the radial pressure and the transverse pressure respectively.  The shape function $b=b(r)$ is used to define the WH spatial shape, more specifically, the throat.  Using the metric singularity we have that
\be
\label{5}
\left. \(1\,-\,\frac{b(r)}{r} \) \right|_{r=r_0} \,=\, 0 \Rightarrow \qquad b(r_{0})=r_0
\ee
where $r_0$ is the throat of the WH.  As a condition for the spacetime be asymptotically flat is $\frac{b(r)}{r} \rightarrow 0$ as $|r| \rightarrow \infty$.  And the shape function follows the conditions: $b'(r_{0}) < 1$ and $b(r) < r \,(r> r_0 )$.

Let us construct the NC version of the shape function.  To carry out this task we will use the idea of Chaichian et al \cite{chaichian} where the authors used the NC underlying relations
\be
\label{6}
[\hat{x}_i,\hat{x}_j ] \,=\,i \theta_{ij}\quad,\quad [\hat{x}_i,\hat{p}_j ] \,=\,i \delta_{ij}\quad,\quad [\hat{p}_i,\hat{p}_j ] \,=\,0\,\,.
\ee

To recover commutativity we can introduce the Bopp shift
\be
\label{7}
x_i\,=\,\hat{x}_i\,+\,\frac 12\,\theta_{ij}\,\hat{p}_j
\ee
where $p_i = \hat{p}_i$ and where the new variables satisfy the usual Poisson brackets,
\be
\label{6.1}
\{ x_i,x_j \} \,=\,0\quad,\quad \{ x_i,p_j \} \,=\,i \delta_{ij}\quad,\quad \{p_i,p_j \} \,=\,0\,\,.
\ee

If we change the variables in (\ref{5}) in order that
\be
\label{8}
1-\frac{b(\hat{r})}{\sqrt{\hat{r}\hat{r}}}\,=\,0
\ee

From (\ref{8}) we can write that,
\ba
\label{9}
&&\frac{1}{\sqrt{\hat{r}\hat{r}}}\,=\,\frac{1}{\sqrt{\(x_i\,-\,\frac 12 \theta_{ij} p_j \)\(x_i\,-\,\frac 12 \theta_{ik}p_k \)}} \nonumber \\
&=&\frac{1}{r}\[1\,+\,\frac{1}{4r^2}\vec{L}\cdot\vec{\theta}\,-\,\frac{1}{8r^2}\(p^2 \theta^2\,-\,(\vec{p}\cdot\vec{\theta})^2 \) \]
\ea
where we have performed an expansion in $\theta$ and we have used only first and second order terms in $\theta$ \cite{chaichian}.  Substituting Eq. (\ref{9}) in Eq. (\ref{8}) we have that
\be
\label{10}
b(r,p) \,=\,r\,-\,\frac{1}{4r}\vec{L}\cdot\vec{\theta}\,-\,\frac{1}{8r}\[p^2 \theta^2\,-\,(\vec{p}\cdot\vec{\theta})^2 \]
\ee

Notice that we wrote $x_i \theta_{ij} p_j = \frac 12 \vec{L} \cdot \vec{\theta}$, where $\vec{L}$ is the throat's angular momentum and $\theta_k = \frac 12 \epsilon_{ijk} \theta_{jk}$.  See \cite{chaichian} for details.  For $\theta \rightarrow 0$ we have $b(r_0)=r_0$, as expected.  

For the equation of state we have that
\be
\label{11}
\omega(\hat{r})\,=\,\frac{b(\hat{r})}{\sqrt{\hat{r}\hat{r}}\,b'(\hat{r})}
\ee
which gives us the following result
\be
\label{12}
\omega(\hat{r})\,=\,\frac{1}{r^2}\left\{r\,-\,\frac{1}{2}\vec{L}\cdot\vec{\theta}\,+\,\frac{1}{8}\[p^2 \theta^2\,-\,(\vec{p}\cdot\vec{\theta})^2 \]\right\}\,\,.
\ee
This last equation is the equation of state for a NCWH.  Notice the presence of angular and linear momenta.

To analyze the energy conditions, from Eqs. (\ref{2}) and (\ref{3}) we have that
\ba
\label{13}
\hat{\rho}\,+\,\hat{p}\,&=&\,\frac{1}{\hat{r}^2}\(b'\,-\,\frac{b}{\hat{r}}\)\nonumber \\
&=&\frac{1}{2 r^2}\left\{\frac{1}{2}\vec{L}\cdot\vec{\theta}\,-\,\frac{1}{8}\[p^2 \theta^2\,-\,(\vec{p}\cdot\vec{\theta})^2 \] \right\} \\
&\times&\[1\,+\,\frac{1}{2r^2}\(\frac 12\,\vec{L}\cdot\vec{\theta}\,-\,\frac{1}{8}\[p^2 \theta^2\,-\,(\vec{p}\cdot\vec{\theta})^2 \] \) \] \nonumber 
\ea
Using the bound for $\theta$ as being $\theta \approx 4,35\,l_{Pl}^2$, where $l_{Pl}$ is the Planck length.  Considering this bound it is easy to see that the angular momentum term is greater than the linear momentum term in (\ref{13}), which has second order $\theta$-terms.  So, both differences encompassing angular momentum in (\ref{13})
are positive and we conclude that $\hat{\rho}\,+\,\hat{p} > 0$.   This result show us that a null geodesic observer would not see negative energy.  In fact, there is no observer passing through a NCWH which is not traversable thanks the result in (\ref{13}).  This last one can also lead us to a possible conclusion: that even at Planck scale, we can have NCWH formed by classical matter which, in general, obeys NEC.  We can ask also if, in this NC regime, the energy-momentum tensor of a scalar field coupled to gravity can violate NEC.  Since we are considering only the signal of these quantities, we will not analyze the components of linear and angular momenta in (\ref{13}) and from now on. 
 
Now let us carry out the summation $\hat{\rho}\,+\,\hat{p}_{tr}$ which is given by
\ba
\label{14}
&&\hat{\rho}+\hat{p}_{tr}\,\,  \\
&=&\frac{1}{r^2}\,
\left\{1\,+\,\frac{1}{4r^2}\,\(\frac 12 \vec{L}\cdot\vec{\theta}\,-\,\frac{1}{8}\[p^2 \theta^2\,-\,(\vec{p}\cdot\vec{\theta})^2 \] \) \right. \nonumber \\
&\times& \left.\[1\,-\,\frac{1}{2r^2}\(\frac 12\,\vec{L}\cdot\vec{\theta}\,-\,\frac{1}{8}\[p^2 \theta^2\,-\,(\vec{p}\cdot\vec{\theta})^2 \] \) \]\right\}  \,>\,0\,\,, \nonumber 
\ea
where we have used the bound for $\theta$.  In Eq. (\ref{14}) the term divided by $2r^2$ is obviously less than one, what makes the difference positive. Hence, from the results obtained in Eqs. (\ref{13}) and (\ref{14}) we can conclude that our NCWH does not violate NEC and from Eq. (\ref{2}) we have that 
\ba
\label{15}
\hat{\rho}\,&=&\,\frac{b'}{8\pi r^2} \nonumber \\
&=&\frac{1}{8\pi r^2} \left\{1\,+\,\frac{1}{2r^2}\[\frac 12\,\vec{L}\cdot\vec{\theta}\,-\,\frac{1}{8}\[p^2 \theta^2\,+\,(\vec{p}\cdot\vec{\theta})^2 \] \] \right\}  \nonumber \\
&&\, >\,0
\ea
and consequently WEC is not violated.  If in this NC regime we had static observers, they would see nonnegative mass-energy density.   It is easy to check from (\ref{15}) that $b' \geq 0$ everywhere.
With the values obtained above it is easy to calculate the NC flaring-out condition and check that,
\be
\label{16}
\frac{\hat{b}\,-\,\hat{b}'\,\hat{r}}{\hat{b}^2}\,>\,0 
\ee
which means that the NCWH can be connected to asymptotically flat spacetime.  Notice that the introduction of noncommutativity via Bopp shift does not modify radically the NCWH spacetime.  Therefore, the possibility that the NCWH flares outward into flat NC spacetime is not an absurd, in our opinion.   The results obtained here means that the noncommutativity introduced new terms in the well known parameters.   On the other hand, of course, it modifies radically the energy conditions considered in the commutative WH scenario, where exotic matter is considered.

%\subsection{}

%\section{}

%\section{Conclusions and perspectives}
Finally, we can say that the NCWHs obey the energy conditions, so we can conclude that in a NC scenario, the construction of a NCWH does not need exotic matter.  Although NEC violation at the throat is a necessary condition to keep the wormhole open, we can ask if there is any amount of exotic matter inside the NCWH in order to make it traversable.  As a future computation we can calculate this minimal amount of energy in order to keep the NCWH throat open and compare with the result in the literature \cite{vkd} for commutative WH.  In other words, the results obtained here does not guarantee that there is any infinitesimal quantity of exotic matter sufficient to keep the throat open.   This is an ongoing research.  In this work we are concerned with the consequences of quantum effects in the NC framework developed here such as the small ANEC violations which, quantically induced, are underlying for the existence of Hawking radiation, for instance.
Therefore, since this exotic matter is responsible for the traversality property of commutative WHs, it is natural to claim that NCWHs do not allow traveling through it.   Besides, since NCWHs do not require exotic matter to exist it seems more arguably to consider the existence of NCWHs than commutative WHs which do need exotic matter.  Namely, of course the WH solution of Einstein equation do exist, however, commutatively speaking, they need a (theoretically) non existing matter to exist as a material cosmological entity.  We saw that, at the Planck scale (NC scale) the NCWHs do not need something to warp spacetime or to provide holes in it.

As future perspectives, since we recognize that the solution of Einstein's equations permit us to elaborate WH physics, a natural way is to add a cosmological constant to the analysis of NCWH framework obtained here.  Hence, in the light of noncommutativity \cite{jhep-nosso} we can ask how to deal with a NCWH formulation of the cosmological constant.  Or if theories yielding bouncing cosmological models \cite{poulis} also generate NCWH solutions.  We can ask also what happens when a NCWH collapses since its commutative partner collapses to become a BH.

In this work we introduced noncommutativity through the Bopp shift.  However, we can imagine the construction of a WH inside a totally NC spacetime, like the one developed in \cite{amorim,abreu} and known as the NC Doplicher-Fredenhagen-Roberts-Amorim spacetime.  With ten dimensions we ask what would be the WH structure within a spacetime where $\theta^{\mu\nu}$ is a coordinate with a conjugate momentum instead of a simple NC parameter, as in the majority NC literature.

It is well known that even below the Planck scale, exists the possibility of other classical violations of NEC, such as higher derivative models, Brans-Dicke theory, or else \cite{bv}.  But all these last ones are based on modifications of general relativity at high energies.
Facing the results obtained here, we believe that an investigation of these issues in a NC geometry background would lead us to interesting paths not explored so far.

\end{document}